# Developing and Building Ontologies in Cyber Security

**Muhammad Shoaib Farooq[1], Muhammad Talha Waseem[1]**
[1]Department of Artificial Intelligence, School of System and Technology, University of Management and Technology, Lahore, 54000, Pakistan
Corresponding author: Muhammad Shoaib Farooq (shoaib.farooq@umt.edu.pk)

**ABSTRACT** Cyber Security is one of the most arising disciplines in our modern society. We work on Cybersecurity domain and in this the topic we chose is Cyber Security Ontologies. In this we gather all latest and previous ontologies and compare them on the basis of different analyzing factors to get best of them. Reason to select this topic is to assemble different ontologies from different era of time. Because, researches that included in this SLR is mostly studied single ontology. If any researcher wants to study ontologies, he has to study every single ontology and select which one is best for his research. So, we assemble different types of ontology and compare them against each other to get best of them. A total 24 papers between years 2010-2020 are carefully selected through systematic process and classified accordingly. Lastly, this SLR have been presented to provide the researchers promising future directions in the domain of cybersecurity ontologies.

## I. INTRODUCTION

Network safety is an arising discipline that pointed toward safeguarding weaknesses or limiting dangers to mechanical foundation like programming, equipment and media transmission. Information about network protection issues is for the most part held by individuals associated with ICT field, because of monstrous utilization of ICT, information about network safety ought to be stretched out to overall population. Information about network safety has been continuously developed thanks to the variety of commitments made by various specialists in this field **[1]**.

A piece of these commitment has centered on making ontologies that assistance to characterize, address and coordinate a jargon of ideas connected with this discipline. These ontologies give shared information on the various perspectives of network protection **[2]**. To have better comprehension of ontologies in the field of network protection, this paper addresses the deliberate survey of the writing on different ontologies **[3] [4]** that have been accounted for with regards to network safety.

In this topic, lots of researches done with respect of time but in most of researches, authors just studied single ontology & explain its pros and cons. We just gather different researches and compare them to get precise and unique research for future researches.

The objective of proposed work is to present a Systematic literature review in the domain of Cyber security ontologies, all we need to assemble all type of ontologies invented in single place. So, we do comparison method with different ontologies with respect to their nature like; Networking, Human factor and more to get valid and desirable set of ontologies in one place.

There are many previous researches on the field of ontologies, Like SENSUS method, Onto Saurus method, Web ODE method all of these methods wander around one specific ontology. But, we assemble all of them here & main reason of this research is that in all previous researches most of them have data before year 2015. So, we gather data after it and compare them with each other to get perfect ontologies of all time. Because, lots of researchers wander here and there but don't get latest set of ontologies to enhance their studies so we give them all a platform to analyze this research and we assure all of them that this is going to be very helpful to them.

The contributions of this paper related to cybersecurity ontologies domain are as follows: In section 2, we represent complete detail about ontologies in the field of cybersecurity. In section 3, we present research methodology by defining research questions, inclusion/exclusion criteria & search string to collect reliable studies related to ontologies in cybersecurity domain. In section 4, we present data analysis by making tables from selected papers. In section 5 & 6 we present conclusions and future directions respectively.

## II. ONTOLOGIES

In the starting points of western contemplations, metaphysics was viewed as a discipline connected with Philosophy. It was situated to the investigation of existing(entities) and their connections. In everyday manner. Metaphysics is characterized a jargon in which substances, gentries, assets, grounds, capabilities & interactions between the components are expressed. Philosophy is significant in light of the fact that it empowers Sharing information about a specific space. In this writing we track down various techniques or ways to deal with make ontologies, for instance: writers propose a methodology for the production of ontologies that support insightful frameworks in view of information.

Additionally, in creators purposed metaphysics creation pointed toward catching information, code it and coming after coordinate it with existing ontologies. Then again, Cactus strategy is introduced. It centers around the development in ideology examining an information base that utilizes a course deliberation, where the setting of the elements is indicated. One of the commonly realized system is Meth-ontology, offered by [1]. It assists with making another cosmology besides junking expiring ones. This system squeezes into an improvement cycle in view of the production of models.

The SENSUS strategy is another methodology, it utilizes existing ontologies making a philosophy skeleton, the subsequent model dispenses with terms inconsequential to the space information. Another philosophy, Onto-Knowledge task that upholds the improvement in ontologies for information the executives. At long last creators notice the KBSI IDEF5 technique, A strategy that permits, assists in creating, change and support of ontologies. Apparatuses additionally important for creating ontologies. Instances of these apparatuses are: Onto lingua host onto saurus, Protégé, Web ODE, Onto Edit with team. Apparatus that is regularly utilized in the formation in ontologies is Protégé is free multiplatform device which is an extensible engineering. Its primary center is the cosmology supervisor, that its usefulness can be reached out using modules.

One more apparatus that upholds the production of ontologies is Onto-saurus. It comprises of two studies: a cosmology host, utilizes an information portrayal framework and an internet browser which permits altering and investigating ontologies utilizing HTML. It's compulsory to take in mind that some of these apparatuses contains unique philosophy improvement language. Example dialects are: XOL (XML-Based Ontology Exchange Language) SHOE (Simple HTML Ontology Extensions) that are augmentation of HTML, DAML+OIL and OWL (Web Ontology Language). Dialect shifts as per their utilization; other than taking into account that dialects would be coordinated with augmentations, APIs laid out by the supplier of dialects. OWL was one of the dialects ordinarily utilized by cosmology engineers. OWL is headed toward distributing and sharing ontologies created on Web. OWL inference of DAML+OIL that distributes a portion of processes.

At the point, Metaphysics is created, a significant perspective to consider that connected with their check and approval (V&V). It very well may be drawn nearer by two viewpoints: concerning their quality with respect to their rightness [2].In the writing we can discover a few methodologies that address the check and approval of a cosmology, for example, the highest quality level methodology [3]; corpus-based; task-based and rules based [1]. As the highest quality level methodology [3], it centers around contrasting the created metaphysics and a reference cosmology made with specific measures. Then again, the corpus-based technique comprises in assessing inclusion of a metaphysics with few ontologies by a corpus in which decided space is fundamentally concealed.

On account of undertaking-based approach, the assessment of a cosmology is focused on a particular errand, in light of the outcomes got to work on the information on this undertaking. One more methodology comprises in approving the philosophy as per a beneficial measure [2], like its construction [1], For instance, quantity hubs that a metaphysics has, is utilized or perplexing models might be utilized. One more methodology depends on specialists, for instance, assessments depend on happenstances of classes, thickness & intervention that nitty gritty in metaphysics. Help instruments for V&V assignments, Onto-Metric, normal language app measurements, Onto-Clean, Eva Lexon and OOPS! [4] For instance, Onto-Clean, permits the assessment of a cosmology in light of its ordered construction. Another one is OOPS! [4], an instrument which checks the cosmology by utilizing a URL to find potential irregularities that might influence its model.

## III. RESEARCH METHODOLOGY

The acknowledgment of a SLR is partitioned into three stages: (1) Planning, (2) Execution and (3) Reporting, following we portray every one of these stages in our unique circumstance. As a feature of the arranging stage, the convention for SLR is created. It settles examination queries & also the targets of this exploration. At first, hotspot looking, search string, and the incorporation and prohibition models are likewise characterized. In the subsequent stage (execution) convention sprints, at this stage we continue with pursuits & records with regard to hunt string characterized in convention, we additionally do the separation of the records as per the consideration and prohibition standards characterized in the convention, we moreover do the examination and union of the pertinent records. At long last, the third stage compares to introduction to finish this SLR.

Principal objective of this SLR is to acquire a superior comprehension in detailed ontologies connected with network protection space. With deference of this goal, following exploration questions have been presented.

### 1. RESEARCH OBJECTIVES (RO)

Main aims and objectives for this research:

RO1: Can we analyze what are the main areas where ontologies actually held.
RO2: Is there any techniques that enhance modern ontologies and make them better from traditional ontologies.
RO3: We need tools that work on techniques for develop better version of ontologies also does we have any.
RO4: All modern ontologies we choose are really validated or just mashup of validation like traditional ontologies.

### 2. QUESTIONS MOTIVATION (QM)

To complete SLR actually, at first, significant exploration questions have been characterized. Further, a far-reaching scan arranging expected survey for distinguishing proof & extraction of articles has been laid out. The exploration assessments tended to in survey with significant inspiration referenced in Table I.

**TABLE I. RQ and their major motivations**

|  | Research Question | Major Motivation |
|---|---|---|
| R-Q1 | Areas of Cyber Security where Ontologies are reported? | To identify specific areas where these ontologies actually held. |
| R-Q2 | Techniques or approaches utilized for advancement of chosen ontologies? | To understand methods that are used for enhancement and improvement of selected ontologies. |
| R-Q3 | Which devices or tools have been utilized for development of announced ontologies? | To understand the various software or tools that are used for development of selected ontologies. |
| R-Q4 | Have revealed ontologies been approved? | To identify weather selected ontologies are validated or not? |

### 3. SEARCHING CRITERIA

Vital period of SLR is readiness of pursuit plan, to find & gather potentially huge articles in picked area satisfactorily. Interaction involves depiction of search string, literature resources used for pursuit, and isolation (inclusion/exclusion) plan to get most significant article's out of assortment. Various parts of articles evaluated subjectively & observationally to address different viewpoints related with examination.

I) SEARCH STRING

A viable & fair examination has directed by figuring out a key-word based string to look & assemble accessible investigations in the field utilizing different notable computerized research repositories. The finished watchwords and their elective terms expected to finalize a quest string for recognizable proof of pertinent articles are determined in Table II. In Table II the '+' sign is utilized for inclusion and '-' sign for exclusion.

**TABLE II. Key-words for search**

| Keywords | Alternate Keywords |
|---|---|
| + Cyber Security Ontologies (CSO) | Ontology Components (OC), Ontologies in Cybersecurity (OCS) |
| + Cybersecurity in ICT (CSICT) | ICT, Cyber Security (CS) |
| - Cybersecurity Vulnerabilities (CSV) | - |

Final search string consists of three sections. Principal piece of string is utilized to restrict outcomes connected with Cyber Security Ontologies and the following part connects with the forecast of its behavior in ICT, while the latter is utilized to keep the outcomes from the inclusion of studies that depend on vulnerability occurs in Cybersecurity. Equation (1) addressing the numerical plan of the search string.

$$R = \forall \left[ (CSO \lor OC \lor OCS) \land (CSICT \lor ICT \lor CS) \not\equiv (CSV) \right] \quad (1)$$

Here in (1), R represents list items get against it, '∀' addressing 'for all', '∨' utilized for 'OR' operator and '∧' for 'AND' operator consolidating with search terms communicated in Table II to make equation of total search string by each selected repository. Nonexclusive inquiry term utilizing (1) can communicated as:

((Cybersecurity Ontologies OR Ontology Components OR Ontologies of Cybersecurity) AND ("Cybersecurity in ICT" OR "ICT" OR "Cybersecurity") NOT (Cybersecurity Vulnerabilities))

II) LITERATURE RESOURCES

Area explicit & unmistakable diaries are chosen to lead writing search from web vaults, devoted research distribution and assortment. The subtleties of chosen vaults, applied search strings and the outcomes are referenced in Table III.

**TABLE III. Publishers' repository wise search strings**

| Repository | Search Strings |
|---|---|
| Science Direct | ((("CYBERSECURITY ONTOLOGIES" OR ONTOLOGY COMPONENTS OR ONTOLOGIES OF CYBERSECURITY) AND (CYBERSECURITY IN ICT OR ICT OR CYBERSECURITY) AND NOT (CYBERSECURITY VULNERABILITIES)) |
| Springer Link | ((CYBERSECURITY ONTOLOGIES OR ONTOLOGY COMPONENTS OR ONTOLOGIES OF CYBERSECURITY) AND ("CYBERSECURITY IN ICT " OR "ICT" OR " CYBERSECURITY ") NOT (CYBERSECURITY VULNERABILITIES)) |

| ACM Digital Library | ((((((''ALL METADATA'':'' CYBERSECURITY ONTOLOGIES '') OR ''ALL METADATA'': ONTOLOGY COMPONENTS) OR ''ALL METADATA'': ONTOLOGIES OF CYBERSECURITY) AND ''ALL METADATA'': CYBERSECURITY IN ICT) OR ''ALL METADATA'': ICT) NOT ''ALL METADATA'': CYBERSECURITY VULNERABILITIES) |
|---|---|
| IEEE Xplore | (((("ONTOLOGIES" OR "ONTOLOGY") AND CYBERSECURITY) OR ONTOLOGY COMPONENTS OR (("CYBERSECURITY" AND "ONTOLOGIES") OR "MACHINE LEARNING")) AND (((("CYBERSECURITY IN ICT" OR "ICT") NOT ("VULNERABILITIES" AND "CYBERSECURITY")))) |

III) INCLUSION & EXCLUSION

Criteria for inclusion criteria (IC) is:
(IC-1) The essential examinations were chosen in view of the title, unique and key-words to decide if they are recognized as important ones, records were chosen considering consistence with the accompanying consideration rules: All papers written in English language.
(IC-2) Papers should be distributed in lofty recorded settings, for example, Diaries, Proceedings and book parts to a companion overview process.
Exclusion criteria applied on those papers that met a portion of the accompanying rejection standards were not considered for this SLR, like
(EC-1) Copied papers, papers whose primary commitment not connected with network protection ontologies, banners, non-English composed papers and short correspondences like letters to manager.
EC-2) If study included vulnerabilities in cybersecurity.

### 4. RELEVANT PAPER SELECTION

At beginning, we select papers from specific time period that resides between 2010 to 2022.When the convention was characterized, we continued with the execution stage. The recently characterized search string is execute for this, utilizing the Scopus data set web search tool. It's compulsory that Scopus is greatest conceptual and reference data set in the field of logical writing. At first 214 report results were acquired subsequent to running the inquiry string. In the wake of applying the consideration and prohibition measures on titles, edited compositions and key-words, 72 reports were chosen

From this choice, the total substance of 69 reports was gotten to. In the wake of breaking down the substance of these documents, 19 were disposed of, so at long last 50 records were considered for this audit (applicable papers). As indicated by recently depicted, we utilized the Scopus data set to run search formula, and items in the applicable records are going by their separate publisher, for example, IEEE, Springer, ACM, Science Direct, among others. Displayed Figure 1, primary Ontology connected with online protection was accounted in 2004 (Simmonds et al., 2004), year onwards, different irregularly work emerges, it's from 2014 that the quantity of ontologies connected with network safety increments, Example, in 2019 eleven ontologies were accounted for (Brita, 2019; Vega Barbas et al., 2019; Gasmi et al., 2019; Danilova et al., 2019; Scarpato et al., 2019; Niyazova et al., 2019; Islam et al., 2019; Basso Moreira et al., 2019; Kats Antonis et al., 2019; Shaaban et al., 2019; Zamfira et al., 2019).

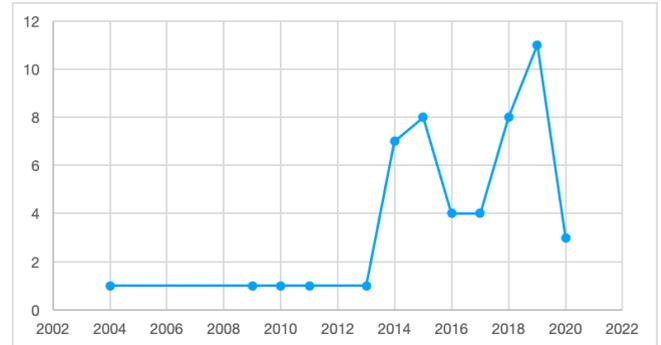

**FIGURE I:** Ontological order revealed with regards to Cybersecurity.

### 5. QUALITY ASSESSMENT CRITERIA

It's compulsory that same papers utilized in SLR are exposed to quality assessment. Nine assessment models adjusted from (Kitchenham) were thought of, these measures are connected with the time of distribution, the sort of distribution (Journal or Conference) also rules connected with the design of the chose reports. A Likert-scale is utilized for this, where greatest assessment score is set 50 focuses, Table IV shows measures utilized for evaluation.

**TABLE IV Evaluation models utilized for this SLR.**

| Sr. | Assessment Questions | Expected Answers | Score |
|---|---|---|---|
| | **Internal Scoring** | | |
| 1 | Is the paper distributed in an effect diary? | a. Yes<br>b. No | a. 5<br>b. 1 |
| 2 | Published time? | a. > 5 year<br>b. < 5 year | a. 5<br>b. 1 |
| 3 | Is the setting of the issue plainly described? | a. Highest<br>b. Lowest | a. 5<br>b. 1 |
| 4 | Is the goal of the study clearly expressed? | a. Highest<br>b. Lowest | a. 5<br>b. 1 |
| 5 | Does the paper contain a result segment? | a. Highest<br>b. Lowest | a. 5<br>b. 1 |
| 6 | Is it conceivable to replicate the review? | a. Highest<br>b. Lowest | a. 5<br>b. 1 |
| 7 | Does the paper address your own research questions? | a. Highest<br>b. Lowest | a. 5<br>b. 1 |
| 8 | Does the review report clear, unambiguous discoveries in an evidence & argument? | a. Highest<br>b. Lowest | a. 5<br>b. 1 |

| 9 | Is the paper well/appropriate referred to? | a. Highest<br>b. Lowest | a. 5<br>b. 1 |

The creators of this paper completed the assessment of the applicable reports and effects of the assessment were arrived to be averaged. As should be visible in Table V, no paper is dismissed in light of the assessment models utilized

**TABLE V: Quality evaluation result.**

| Quality % by classification | No of articles | % from articles |
|---|---|---|
| Worst (<26%) | 0 | NIL |
| Average (26%-45%) | 0 | NIL |
| Better (46%-65%) | 19 | 38% |
| Best (66%-85%) | 26 | 52% |
| Extraordinary (>86%) | 5 | 10% |

The complete score from each paper is figured utilizing a rate. We see every one of important papers yielded a quality score, that reaches from great to phenomenal. Typical score of assessment is 76%, which is viewed as sufficient quality marker for this SLR.

## IV. DATA ANALYSIS

Section gathers results & gives sophisticated answer of every chose paper. Chose papers are explored to successfully respond to the research questions. Initial segment of this part talks about the query items acquired by characterized search formula. Portrayal of appraisal score and at last is committed to exhaustive conversations respond to research assessments.

### 1. SEARCH RESULTS

The essential pursuit process yields a complete 553 articles from different web-based information sources. On this assortment, choice cycle portrayed in the past area was applied. The stages associated with choice cycle have additionally been portrayed beneath Figure 2 & stage wise determination results are communicated in Table VI.

**TABLE VI. Repository based selection**

| Repository | Primary Search | P-I | P-II | P-III |
|---|---|---|---|---|
| IEEE XPLORE | 233 | 74 | 58 | 11 |
| SCIENCE DIRECT | 170 | 68 | 32 | 8 |
| SPRINGER LINK | 110 | 55 | 27 | 3 |
| ACM DIGITAL LIBRARY | 40 | 13 | 7 | 2 |
| TOTAL | 553 | 210 | 124 | 24 |

Title based choice is perform by two writers at stage I (P-1), brings about the determination of 210 papers. Then, copy papers are eliminated in stage (P-II), and space immaterial papers are likewise screen based on consideration and rejection measures characterized in past segment. Unique based screening is applied in stage III (P-III) on the 124 resultant papers acquired from past stage & absolute 24 articles were seen as generally relevant and settled to remember for this SLR for information extraction and examination.

Very acknowledged advanced libraries (DL) to distribute research reads up for different Journals, Conference & Workshops were utilized to choose reads up for this methodical writing survey according to look through methodology displayed in Table III. DL-wise dispersion proportion of chosen articles, incorporates ACM Digital Library 6%, 56% of IEEE Xplore and Science Direct 24%, and 14% of Springer Link. Distributer based organized savvy choice status and appropriation proportion of chosen examinations has previously been displayed in Table V.

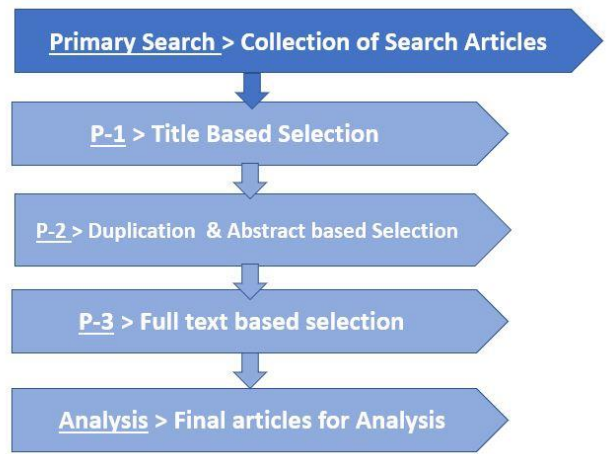

**FIGURE 2. Selection procedure**

### 3. RESEARCH QUESTIONS ASSESSMENT

The 24 primary articles were examined based on research questions portrayed in Table I. The realities extricated after examination of chose studies are based on examens the assessment of data.

1) QUESTION 1'S ASSESSMENT. AREAS OF CYBERSECURITY WHERE ONTOLOGIES ARE REPORTED?

Concerning question, we distinguished 4 classes where ontologies are assembled: General, Networking, Software & Human Factor. Overall class alludes for ontologies that include a blend of ideas connected with different classes, for example, systems administration, programming, or the human variable. In the class of system' administration the ontologies address ideas connected with gear, conventions and organization displaying. In the product classification, the ontologies are chiefly centered around portraying online protection ideas from a product improvement point of view. In the classification of human element are those ontologies that portray ideas connected with the staff engaged with parts of online protection in ICTs. Table 7 presents the ontologies gathered by the classifications recently portrayed. As displayed in Table 7, biggest number of ontologies broke down are gathered in general and systems administration classifications.

**TABLE VII. Classification of ontologies according to their scope.**

| SCOPE | No of Papers | % of Papers | References |
|---|---|---|---|
| General | 9 | 34% | [1]Petrenko, S. A., & Makoveichuk, K. A [2] Petrenko, S. A., & Makoveichuk, K. A [3] Petrenko, S. A., & Makoveichuk, K. A |
| Networking | 8 | 32% | [4] Chukkapalli et al. (2020), Scarpato et al. (2019), Katsantonis et al. (2019), [3] Zamfira et al. (2019), Zamfira et al. (2018), Mozzaquatro et al. (2018), [5] Zheng et al. (2018), Albalushi et al. (2018), Bergner & Lechner (2017), |
| Software | 4 | 18% | [6] Syed (2020), Bataityte et al. (2020), Gasmi et al. (2019), [7] Alqahtani & Rilling (2017), |
| Human Factor | 3 | 16% | [8] Takahashi and Kadobayashi (2014), Takahashi & Kadobayashi (2011) |

### 2) QUESTION 2'S ASSESSMENT. WHICH TECHNIQUES OR APPROACHES UTILIZED FOR ADVANCEMENT OF CHOSE ONTOLOGIES?

With respect to investigate question, we see that the greater part of the creators of the revealed ontologies (88%,46 papers) don't specify to involve existing techniques for the improvement of their ontologies. All in all, the creators portray their own methodology that they followed for the advancement of their ontologies. Less significantly we see that main in seven pertinent papers (12%), the creators notice following some current philosophy for the advancement of their ontologies. For instance, four creators notice the utilization of the Meth-ontology technique (Fernandez et al.), while two creators notice the utilization of the system Ontology Development 101 (Noy & McGuinness).

**TABLE VIII. Ontologies classification according to their Methodology.**

| EXISTING METHODS | NO OF PAPERS | % OF PAPERS | REFERENCES |
|---|---|---|---|
| Without following existing methods | 17 | 88% | [4] Chukkapalli et al. (2020), [9] Mozzaquatro et al. (2018), [7] Alqahtani and Rilling (2017), [1] Petrenko and Makoveichuk (2017), [8] Takahashi and Kadobayashi (2014), Takahashi and Kadobayashi,(2011), Takahashi et al. 2010), Hieb et al., Simmonds et al. |
| Following existed methods | 2 | 12% | [10] Kats Antonis et al. (2019), [11] Van Vuuren et al. (2014) |

As displayed in Table 8, just seven papers report the utilization of a system for fostering ontologies. Meth-ontology approach was accounted in 4 papers (Zamfira et al., 2019; Matheus et al., 2018; Obrest et al., 2014; Razzaq et al., 2014), though the philosophy improvement 101 technique is accounted in two works (Kats Antonis et al.,2019; Van Vuuren et al., 2014).

### 3) QUESTION 3'S ASSESSMENT. WHICH DEVICES OR TOOLS ARE UTILIZED FOR DEVELOPMENT OF ANNOUNCED ONTOLOGIES?

On account of the instruments that help the development of ontologies, 40% of the reports (20 papers) creators notice the utilization of a device to help the improvement of their ontologies. We see that Protégé is the most involved instrument for philosophy improvement; its utilization is referenced in 16 out of 20 significant reports that notice the utilization of a device. Less significantly, the utilization of different devices is additionally revealed, for example, Atom-Tool; Cyber Security Ontology Expert Tool; CYBEX; and Intel-MQ.

**TABLE IX. Ontologies arrangement by utilized devices**

| TOOL | NO OF PAPERS | % OF PAPERS | REFERENCES |
|---|---|---|---|
| | | | [10] Katsantonis et al. (2019), [9] Mozzaquatro et al. (2018), |

| | | | [8] Razzaq et al. (2014) |
|---|---|---|---|
| Protégé | 16 | 68% | |
| **ATOM-TOOL** | 2 | 8% | [12] Moreira (2018) |
| **Cyber Security Ontology Expert** | 2 | 8% | [11] Singer. P (2014) |
| **CYBEX** | 2 | 8% | [8] Takahashi and Kobayashi (2014) |
| **INTEL-MQ** | 2 | 8% | [13] Zheng. H |

The apparatuses utilized for philosophy improvement typically consolidate dialects that assistance in meaning of ontologies. We see that 24% (12 papers) of pertinent archives (Chukka Alli et al., 2020; Vega Barbas et al., 2019; Doynikova et al., 2019; Zheng et al., 2018; Petrenko and Makoveichuk, 2017; Elnagdy et al., 2016; Falk, 2016; Syed et al., 2016; Iannacone et al., 2015; Oltramari et al., 2015; Salem and Wacek, 2015; Laskey et al., 2015) report the utilization of the OWL language (Dean & Schreiber).

Significantly, utilization of dialects, for example, SPARQL, SWRL, XML and OWL 2 is noticed (Baesso Moreira et al., 2019; Onwubiko et al., 2018; Albalushi et al., 2018; Tseng et al., 2017; Bergner and Lechner, 2017; Fontenele and Sun, 2016; Maines et al., 2015; Geller et al., 2014).

4) QUESTION 4'S ASSESSMENT. HAVE REVEALED ONTOLOGIES BEEN APPROVED?

As to assessment and approval of ontologies revealed in SLR, 62% of them (31 essential examinations) notice utilization of confirmation or approval systems. Instance, 18 applicable articles, creators notice utilization of data extraction rules like those proposed in (Boley et al.). On account of the ontologies here detailed, a few creators play out the check and approval of their ontologies in view of the correlation with existing ontologies (3 significant papers). One more kind of approval noticed is the approval by specialists, move toward referenced in other three important papers. The utilization of devices for evaluating ontologies is additionally referenced, instruments like Onto-Clean (Gangemi et al.), the OQuare measurements device (Duque and Fernandez, 2011), Protégé expansion called HermiT Reasoner (Data and Knowledge Group) are referenced. We additionally notice half and half methodologies in which approval is led using models (Fernandez et al., 2009) and errands (Welty et al.). At long last, we likewise notice the utilization of a measurement base approval approach
In Table X, all of them are briefly explained:

**TABLE X. Ranks of publication sources of selected articles**

| EVALUATION | No of Papers | % of Papers | REFERENCES |
|---|---|---|---|
| **Information Extraction Rules** | 11 | 66% | [13] Zheng. H (2018)<br><br>[7] Alqahtani and Rilling (2017)<br><br>[1] Petrenko an Malkovich (2017)<br><br>[8] Takahashi and Kad obayashi (2017), Takahashi and Kad Obayashi (2011)) |
| **Onto-Clean** | 3 | 6% | [14] Read. J & Cruz. C |
| **Expert's Validation** | 3 | 6% | [15] Duque-Ramos (2011) |
| **Comparison by existing *ontology*** | 3 | 6% | [13] Zheng et al. (2018) |
| **Hermit Reasoner** | 1 | 4% | [16] Niyazova (2011) |
| **O-*Quare Metrices*** | 1 | 4% | [9] Mozzaquatro et al. (2018) |
| **M*etrice Base Validation*** | 1 | 4% | [17] Scarpato (2019) |
| **Hybrid Approch** | 1 | 4% | [18] Ulanov (2010) |

## V. DISCUSSIONS & CONCLUSION

The most elevated level of ontologies detailed in the field of online protection are in the general and systems administration classification. The overall class tends to a blend of ideas having a place with the systems administration, programming and human variable classes. These discoveries propose that work is being finished on the meaning of ontologies in unambiguous network protection areas. Notwithstanding, we likewise notice work on the improvement of ontologies tending to broad space of the network protection.

As to strategies or approaches utilized for the improvement of ontologies, we see that in the majority of the important reports examined, creators don't make reference to following existing philosophies for fostering their ontologies. Just in 12% creators notice involving some procedure as a kind of perspective. From this rate, the most utilized procedure is Meth-ontology (Fernandez et al.). These discoveries appear to recommend an absence of inspiration in the utilization of existing metaphysics improvement techniques. We see that the most generally involved apparatus for development of ontologies is Protégé (Noy et al.).

From ontologies that report utilization of some device, 32% report utilization of Protégé. OWL (Dean & Schreiber) is most utilized language among ontologies that report utilization of language.

At last, regards to utilization of confirmation and approval (V&V) components, we see that 62% of essential examinations report utilization of check or approval system, with accompanying stick out: utilization of data extraction rules, making of correlations as for existing ontologies, approval by specialists, approval using measurements, also utilization of apparatuses for this. We see that continuously there is interest in applying V&V components that assistance address lacks in advancement of ontologies in area of Cyber security.

## VI. FUTURE DIRECTIONS

It's compulsory that optional examinations as detailed are dependent upon understanding in various stages, in this manner suggesting the presence of predisposition. To limit a potential predisposition, we followed the principal stages with exercises of SLR philosophy. Albeit, gamble of missing significant papers was available, think about that, chose records for this survey (pertinent papers) address a sufficient example of revealed online protection ontologies.

In this SLR, we didn't think about dim writing, so we accept that great quality dark writing of this subject will be accounted for in diaries or meetings, along these lines, conceivable distribution predisposition might emerge because of negative discoveries are not normally distributed. We didn't consider reports distributed not in English language, albeit isn't a restriction in our provincial setting, it tends to be an impression of the limits forced on us by the accessible examination around here (refreshed and peer-evaluated writing is regularly distributed in English). Network safety is a youthful discipline wherein trains, for example, media communications, hardware and processing have met. The appearance of changed ontologies in the field of network protection has cultivated more information on ideas connected with this discipline. In this examination we have detailed consequences of deliberate writing survey on network protection ontologies. The primary commitment of this auxiliary review is the blend of the discoveries from the various ontologies announced concerning regions or spaces, systems, instruments and dialects utilized, as well as the V&V components revealed. The consequences of our work can act as a kind of perspective for future examination on this subject.